\documentclass[reprint,amsmath,amssymb,aps,superscriptaddress,prb,floatfix]{revtex4-2}
    \usepackage{graphicx}           
    \usepackage{dcolumn}            
    \usepackage{bm}                 
    \usepackage{hyperref}           
    \hypersetup{colorlinks=true, linkcolor=blue, citecolor=blue, urlcolor=blue, pdftitle={Experimental observation of metallic states with different dimensionality in a quasi-1D charge density wave compound}, pdfauthor={PedroRezende}}
    \usepackage{orcidlink}
    
\begin{document}
\preprint{APS/123-QED}

\title{Experimental observation of metallic states with different dimensionality in a quasi-1D charge density wave compound}

\author{P. Rezende-Gon\c calves\,\orcidlink{0000-0001-6773-8921}}
    \email{pedro.rezende-goncalves@universite-paris-saclay.fr}
    \affiliation{Universit\'{e} Paris-Saclay, CNRS, Institut des Sciences Mol\'{e}culaires d'Orsay, 91405 Orsay, France}
    \affiliation{Departamento de F\'{i}sica, Universidade Federal de Minas Gerais, Av. Pres. Antonio Carlos, 6627 Belo Horizonte, Brazil}
\author{M. Thees\,\orcidlink{0000-0002-6527-0373}}
    \affiliation{Universit\'{e} Paris-Saclay, CNRS, Institut des Sciences Mol\'{e}culaires d'Orsay, 91405 Orsay, France}
\author{J. Rojas Castillo}
\author{D. Silvera-Vega}
    \affiliation{Department of Physics, Universidad de Los Andes, Bogot\'{a} 111711, Colombia}
\author{R. L. Bouwmeester\,\orcidlink{0000-0003-1114-5152}}
    \affiliation{Faculty of Science and Technology and MESA+ Institute for Nanotechnology, University of Twente, 7500 AE Enschede, Netherlands}
\author{E. David\,\orcidlink{0000-0002-4406-7501}}
\author{A. Antezak}
\author{A. J. Thakur}
\author{F. Fortuna\,\orcidlink{0000-0002-7063-1890}}
    \affiliation{Universit\'{e} Paris-Saclay, CNRS, Institut des Sciences Mol\'{e}culaires d'Orsay, 91405 Orsay, France}
\author{P. Le F\`{e}vre\,\orcidlink{0000-0001-9800-8059}}
    \affiliation{Synchrotron SOLEIL, L'Orme des Merisiers, Saint-Aubin-BP48, 91192 Gif-sur-Yvette, France}
\author{M. Rosmus\,\orcidlink{0000-0002-4314-9601}}
\author{N. Olszowska\,\orcidlink{0000-0001-8377-4615}}
    \affiliation{National Synchrotron Radiation Centre SOLARIS, Jagiellonian University, Czerwone Maki 98, 30-392 Krak\'{o}w, Poland}
\author{R. Magalh\~{a}es-Paniago}
    \affiliation{Departamento de F\'{i}sica, Universidade Federal de Minas Gerais, Av. Pres. Antonio Carlos, 6627 Belo Horizonte, Brazil}
\author{A. C. Garcia-Castro\,\orcidlink{0000-0003-3379-4495}}
    \affiliation{School of Physics, Universidad Industrial de Santander, Carrera 27 Calle 09, 680002, Bucaramanga, Colombia}
\author{P. Giraldo-Gallo\,\orcidlink{0000-0002-2482-7112}}
    \affiliation{Department of Physics, Universidad de Los Andes, Bogot\'{a} 111711, Colombia}
\author{E. Frantzeskakis\,\orcidlink{0000-0002-6014-5637}}
    \email{emmanouil.frantzeskakis@universite-paris-saclay.fr}
\author{A. F. Santander-Syro\,\orcidlink{0000-0003-3966-2485}}
    \email{andres.santander-syro@universite-paris-saclay.fr}
    \affiliation{Universit\'{e} Paris-Saclay, CNRS, Institut des Sciences Mol\'{e}culaires d'Orsay, 91405 Orsay, France}
\begin{abstract}
TaTe$_4$ is a quasi-1D tetrachalcogenide that exhibits a CDW instability caused by a periodic lattice distortion. Recently, pressure-induced superconductivity has been achieved in this compound, revealing a competition between these different ground states and making TaTe$_4$ very interesting for fundamental studies. Although TaTe$_4$ exhibits CDW ordering below 475~K, transport experiments have reported metallic behavior with a resistivity plateau at temperatures lower than 10~K. In this paper, we study the electronic structure of TaTe$_4$ using a combination of high-resolution angle-resolved photoemission spectroscopy and density functional calculations. Our results reveal the existence of the long-sought metallic states. These states exhibit mixed dimensionality, while some of them might have potential topological properties.
\end{abstract}

\maketitle

\section{Introduction}
Since the discovery of new topological phases of matter in the last decade \cite{thouless1982,moore2010}, the general classification of the electronic structure of solids has been largely enriched. Topological insulators \cite{hasan2010,hsieh2008,xia2009}, superconductors \cite{bednorz1986,damascelli2003}, and an extended set of remarkable materials \cite{burkov2016,sato2017,dzero2016,sekine2021} have made condensed matter physics much more complex. Because of their technological potential \cite{novoselov2012,he2019,kruglyak2010,prinz1999}, the investigation of novel materials is crucial \cite{cohen2016}. In this context, transition metal chalcogenides (TMCs) stand out for their diversity and versatility \cite{mitchell2002,manzeli2017}. The binary combination of a transition metal and an element from the chalcogen family (S, Se or Te) produces, in most cases, a layered compound that foments the appearance of interesting physical properties related to low dimensionality \cite{splendiani2010,wang2017}.

A notable example of materials with intriguing properties induced by low-dimensionality is the family of quasi-1D transition metal chalcogenides. These compounds are known for their crystalline structure formed by linear chains of transition metal atoms surrounded by chalcogen atoms \cite{patra2020,island2017,gressier1984}. This peculiar arrangement has important implications for the electronic transport in those compounds, producing effects that should be observed in a theoretical one-dimensional crystal, such as charge density waves (CDWs) \cite{gruner1988,gruner1994,gorcov1989}. This is the case for the tellurides NbTe$_4$ and TaTe$_4$, both of which exhibit charge density ordering below room temperature \cite{boswell1983,galvis2023}. Recently, pressure-induced superconductivity has also been reported for both compounds \cite{yang2018,yuan2020}, demonstrating the interplay between these two non-trivial ground states.

Furthermore, TaTe$_4$ has been predicted to exhibit topological properties related to the existence of degeneracy points in its band structure, referred to as Dirac nodes \cite{zhang2020}. In addition, magnetotransport experiments have revealed the metallic behavior of TaTe$_4$ with a field-induced metal-to-insulator transition at about 35~K, followed by a resistivity plateau below 10~K \cite{luo2017,gao2017}. In non-magnetic systems \cite{pickem2021}, a resistivity plateau can be a fingerprint of low-dimensional states, including topological surface states, where the presence of protected metallic surface states saturates the bulk insulating resistivity at low temperatures \cite{ren2010,kim2014}. The same feature is observed in topological semimetals when exposed to magnetic fields \cite{sun2016,wang2016}, but the underlying mechanism is still under debate.

The coexistence of 1D and 3D electronic bands in TaTe$_4$ has been revealed more than twenty years ago through angle-resolved photoemission spectroscopy (ARPES) measurements, albeit no metallic states (i.e., no states crossing the Fermi level)  were observed \cite{zwick1999}. Theoretically, in quasi-1D materials, the photoemission intensity vanishes at the chemical potential and no Fermi level cutoff is expected \cite{dardel1991,hwu1992}. In contrast, TaTe$_4$ exhibits a well-defined Fermi level cutoff within a region of reduced spectral weight ranging from the Fermi level down to about $-0.2$~eV. Although optical conductivity measurements \cite{zwick1999} indicate the presence of free carriers near the Fermi level, in addition to the metallic character revealed by transport measurements, no dispersive states have been observed in this region. The apparent inconsistency regarding the existence and nature of metallic states in TaTe$_4$ calls for a new experimental investigation of its band structure.

In the present work we provide new insights into the electronic bandstructure of TaTe$_4$ by means of high-resolution ARPES and density functional calculations (DFT). We experimentally probe the long-sought metallic states that cross the Fermi level. These states form 3D Fermi surface contours which are accompanied by quasi-1D Fermi contours that connect points of band intersections.

\section{Materials and Methods}
\subsection{Sample Growth}
Single crystals of TaTe$_4$ were grown by the self flux method \cite{fisher2012}. Alumina crucibles containing mixtures of 1 mol \% elemental Ta and 99 mol \% elemental Te powders were placed in an evacuated, sealed quartz tube. The mixture was heated in a box furnace up to 700$^{\circ}$C and kept at this temperature for 12 hours. The temperature was then gradually decreased to 500$^{\circ}$C at a rate of 2$^{\circ}$C/hour. The containers were quickly transferred to a centrifuge, where the TaTe$_4$ crystals were separated from the remaining melt. Once at room temperature, silver-colored rectangular crystals with dimensions up to 0.1$\times$0.1$\times$1~cm$^3$ were obtained. The formation of a single-crystalline phase was confirmed by X-ray diffraction.

\subsection{Experimental Methods}
ARPES measurements were carried out using hemispherical electron analyzers with vertical slits at the CASSIOPEE beamline of Synchrotron SOLEIL (France) and the URANOS beamline of Synchrotron SOLARIS (Poland). In order to generate pristine surfaces, TaTe$_4$ crystals were cleaved in-situ, exposing the (100) crystalline plane. Typical resolutions in electron energy and angle were 15~meV and 0.25$^{\circ}$, respectively, in both experimental setups. ARPES measurements were performed at temperatures not higher than 20~K and pressure below 5$\times$10$^{-11}$~mbar. Measurements with a photon energy of 75~eV correspond to a measurement plane right across the center of the bulk Brillouin zone [i.e. blue plane highlighted in Fig. 1(b)] assuming an inner potential of 18~eV.

\subsection{Computational Approach}
Theoretical analyses were performed within the framework of first-principles calculations in the density-functional theory (DFT) \cite{hohenberg1964,kohn1965} approach. The mentioned calculations were  carried out in the Vienna $ab$-initio simulation package, \textsc{vasp} code (version 5.4.4) \cite{kresse1996,kresse1999}. The projected-augmented waves approach, PAW \cite{blochl1994}, was employed to represent the valence and core electrons. The electronic configurations considered in the pseudo-potentials as valence electrons are Ta: (5$p^6$6$s^2$5$d^3$, version 07Sep2000) and Te: (5$s^2$5$p^4$, version 08April2002). The exchange-correlation was represented within the generalized gradient approximation (GGA-PBEsol) parametrization \cite{perdew2008}. The periodic solution of the crystal was represented by using Bloch states with a Monkhorst-Pack \cite{monkhorst1976} \emph{k}-point mesh of 13$\times$13$\times$13 in the high-symmetry $P4/mcc$ (SG. 124) phase and scaled to 9$\times$9$\times$5 in the low-symmetry CDW $P4/ncc$ (SG. 130) phase into the $\sqrt{2}$$\times$$\sqrt{2}$$\times$3 supercell representation. We used a 600~eV energy cut-off to ensure forces convergence of less than 0.001~eV$\cdot$\r{A}$^{-1}$ and energy less than 0.1~meV. Spin-orbit coupling (SOC) effect was considered as implemented in the \textsc{vasp} code \cite{hobbs2000}. To analyze the electronic structure, we made use of the \textsc{Python} library \textsc{PyProcar} \cite{herath2020}. Finally, the structural figures illustrations were performed with the \textsc{vesta} code \cite{momma2011}.

\section{Results and Discussions}
\begin{figure*}
    \centering
    \includegraphics[width=1\textwidth]{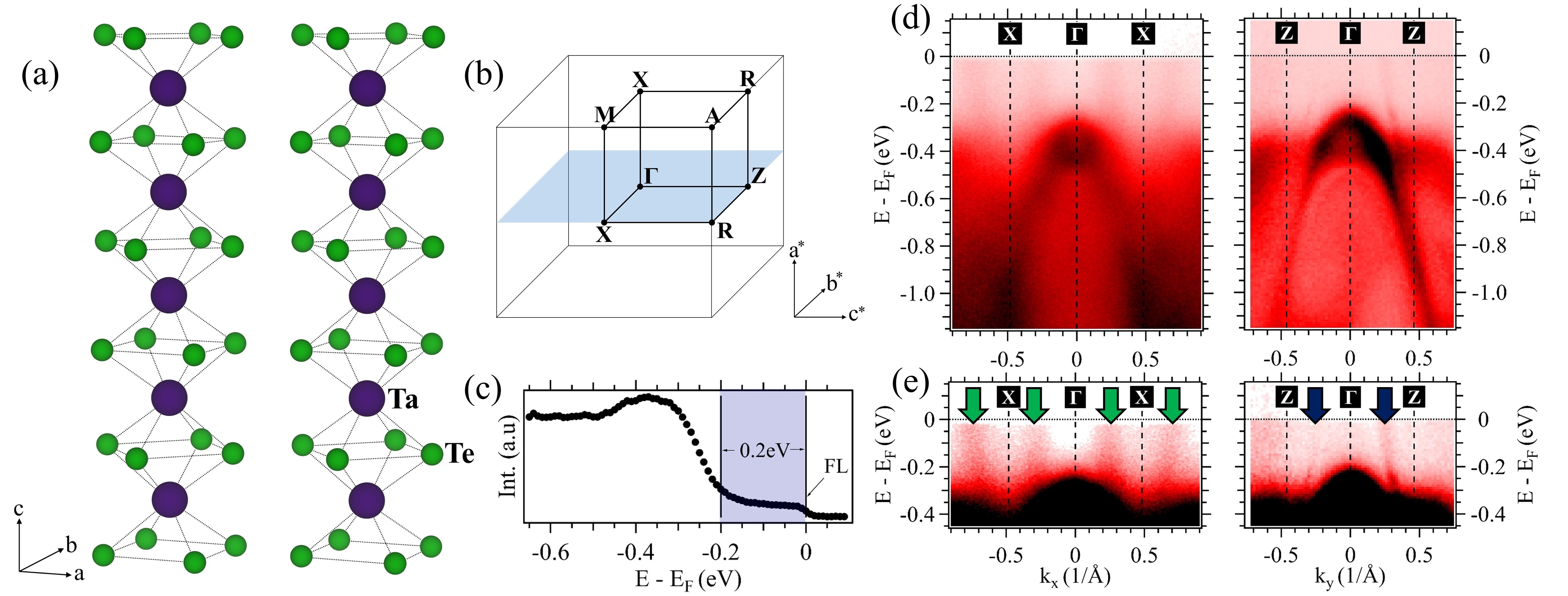}
    \caption{(Color online) (a) Crystal structure of TaTe$_4$ showing the Ta linear chains and the regular octahedra formed by Ta and the Te squares. Ta and Te atoms are represented by purple and green spheres, respectively. (b) Reciprocal unit cell of TaTe$_4$ and the corresponding high-symmetry points. The measurement plane perpendicular to the $a^*$-axis is highlighted. (c) Angle-integrated intensity map of TaTe$_4$ showing the Fermi-level cutoff and the suppression of spectral intensity in an energy window of 0.2~eV. (d) ARPES energy-momentum maps of TaTe$_4$ along $\Gamma$-X and $\Gamma$-Z, showing the dispersion perpendicular and parallel to the linear chains, respectively. (e) Energy-momentum maps along $\Gamma$-X and $\Gamma$-Z with saturated contrast over the region of reduced spectral intensity, showing electron-like metallic states (indicated by green arrows) and a hole-like metallic state (indicated by blue arrows) crossing the Fermi-level. The Brillouin zone boundaries are marked by dashed lines in all ARPES images. All measurements were carried out with a photon energy of 75~eV and linear-horizontal (LH) polarization.}
    \label{FIG1}
\end{figure*}

The crystal structure of TaTe$_4$ \cite{bjerkelund1964} consists of linear chains of Ta atoms surrounded by Te atoms, with each atom representing a vertex of a regular octahedron. The underlying square bases are rotated by 45$^{\circ}$ with respect to each other, resulting in an antiprismatic configuration [Figure \ref{FIG1}(a)]. The high symmetry (in the non-CDW phase) unit cell is the tetragonal $P4/mcc$ (SG 124) with lattice parameters: $a = b =$ 6.514~\r{A} and $c =$ 6.809~\r{A}, where the major axis is oriented parallel to the chains. The reciprocal unit cell is shown in Figure \ref{FIG1}(b). The natural cleavage plane is perpendicular to $a$ (or equivalently $b$), hence the measurements plane is defined by the $b$ and $c$ axes (or equivalenty $a$ and $c$ axes). Lattice distortions in TaTe$_4$, driven by a V$_4$ mode as in the NbTe$_4$ case \cite{galvis2023}, are responsible for a CDW transition at $T = 475$~K that leads the system into a commensurate 2$\times$2$\times$3 $P4/ncc$ structure \cite{boswell1983}. Although the atomic structure of TaTe$_4$ has been known for many years, its electronic band structure is still not fully understood.

\subsection{Electronic band structure by ARPES}
According to the electronic configuration of Ta and Te atoms, the unpaired electrons must be located in the Ta-5$d$ and Te-5$p$ shells, and their hybridization results in the band structure of TaTe$_4$ near the Fermi level \cite{guster2022}. In-plane ARPES measurements show that the electronic structure near the Fermi level consists of a hole-like band centered at the $\Gamma$-point and an electron-like band centered the X-point [Fig. \ref{FIG1}(d)]. The spectral intensity is significantly reduced near the Fermi level, a particular interesting character of TaTe$_4$, which might be related to the CDW gap opening. The width of the region of reduced spectral weight is around 0.2~eV [Fig. \ref{FIG1}(c)]. In contrast to Zwick and coworkers \cite{zwick1999}, our measurements distinctly show metallic states crossing the Fermi level --see Figure \ref{FIG1}(e). The observation of these states in the aforementioned region of reduced spectral weight solves the alleged mystery of TaTe$_4$ exhibiting metal-like resistivity curves \cite{luo2017} without any, so far, experimental fingerprint of a metallic state \cite{zwick1999}.

\begin{figure*}
    \centering
    \includegraphics[width=1\textwidth]{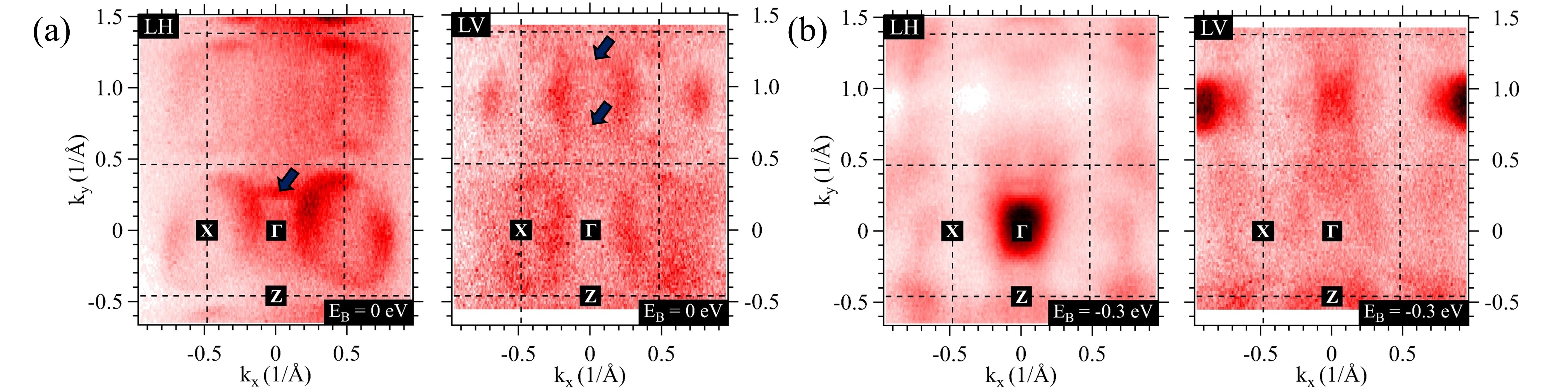}
    \caption{(Color online) (a) ARPES in-plane Fermi surface maps ($E_B=0$) of TaTe$_4$ obtained with photons of linear horizontal (LH, left panel) and linear vertical (LV, right panel) polarization, showing the in-plane contours of the electron-like metallic states (elliptical contours) and of the quasi-1D metallic states (indicated by arrows). (b) ARPES constant energy maps of TaTe$_4$ at $E_B=-0.3$~eV obtained in both polarizations (LH on the left and LV on the right), showing the contours at the band maximum of the hole-like states centered at $\Gamma$. The Brillouin zone boundaries are marked by dashed lines in all images. All measurements were carried with a photon energy of 75~eV.}
    \label{FIG2}
\end{figure*}

The electron-like band around X gives rise to elliptical electron pockets that are the most readily observable features of the Fermi surface contours. The corresponding Fermi surface maps shown in Figure \ref{FIG2}(a) are obtained using two photon polarizations: linear horizontal (LH) and linear vertical (LV). The spectral weight of the Fermi contours is strongly influenced by the crystal orientation, as manifested by the lack of symmetry between adjacent Brillouin zones, and by the polarization of incoming photons. This influence is due to photoemission matrix elements, which are responsible for enhancing or attenuating the spectral signal of different wavefunctions \cite{moser2017,santander2011,rodel2016}. The dispersion of the metallic states is reminiscent of the Dirac-like dispersion of surface states reported for several materials, including topological insulators \cite{zhang2009,goncalves2019,arakane2012}. Nevertheless, before drawing any conclusion on the origin of these states, their dimensionality will be further discussed in a later section by means of photon-energy-dependent ARPES and through comparison to DFT calculations. We note that the lower-lying electronic states of TaTe$_4$ are also subject to strong photoemission matrix elements effects as revealed by the contours of the band maximum of the hole-like state centered at the $\Gamma$-point [Fig. \ref{FIG2}(b)].

We now draw the readers' attention to the hole-like metallic state crossing the Fermi level along $\Gamma$-Z [Fig. \ref{FIG1}(e), right panel]. This state forms quasi-1D Fermi surface contours pointed out by arrows in Figure \ref{FIG2}(a). This means that it poorly disperses along any direction normal to the $\Gamma$-Z high-symmetry line. In terms of spectral intensity, the most intense quasi-1D feature is observed at $k_y=0.27$~\r{A}$^{-1}$ in the LH Fermi surface map, while weaker features are found at $k_y=0.66$~\r{A}$^{-1}$ and $k_y=1.17$~\r{A}$^{-1}$ in the LV Fermi surface map. The asymmetric intensity distribution observed on different in-plane Fermi surface maps is again a manifestation of the photoemission matrix elements \cite{moser2017,santander2011,rodel2016}. In the following section, we will further discuss the electronic structure of TaTe$_4$ through a comparison of the experimental results with DFT calculations.

\subsection{Electronic band structure by DFT}
\begin{figure*}
    \centering
    \includegraphics[width=1\textwidth]{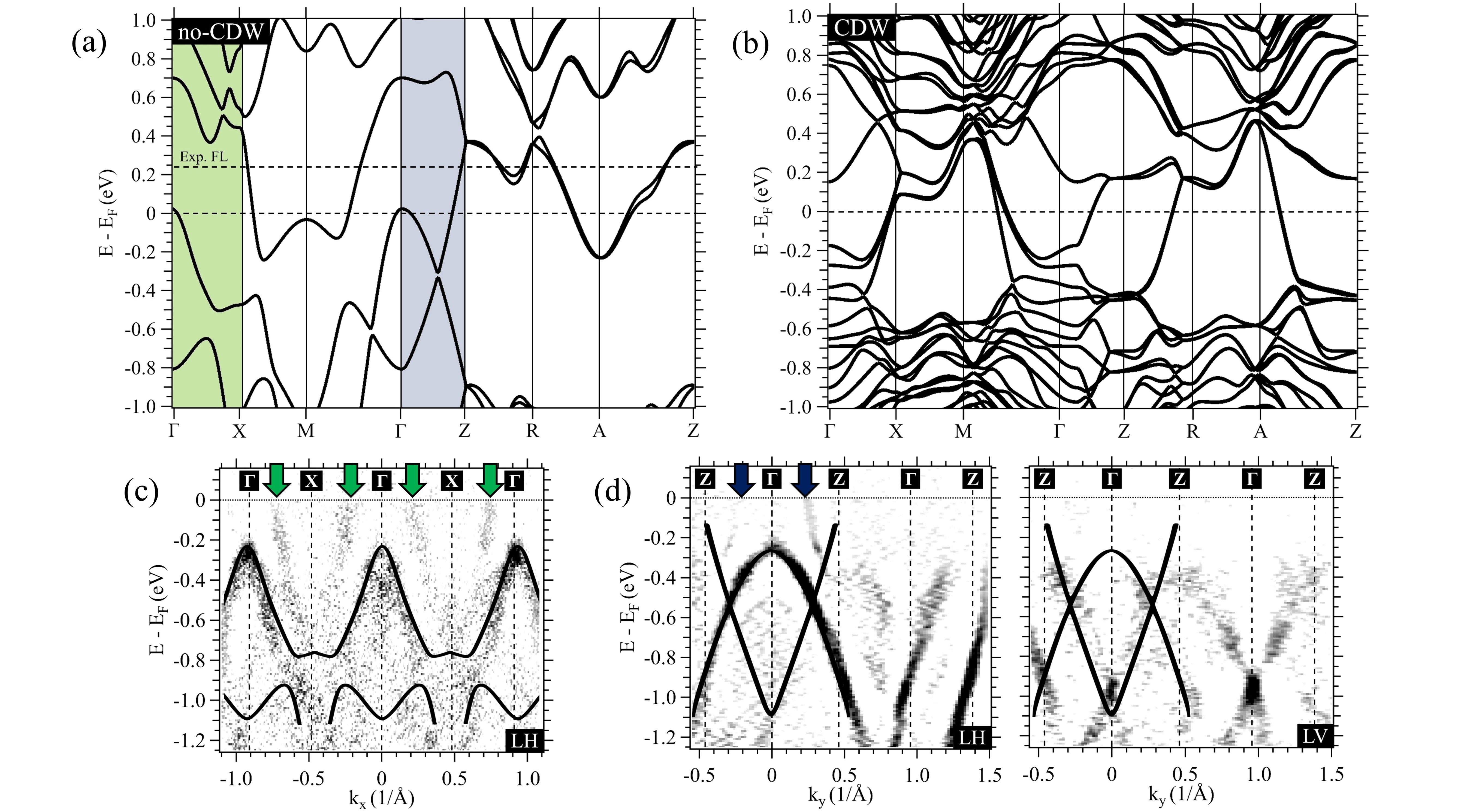}
    \caption{(Color online) (a) Electronic bandstructure of TaTe$_4$ calculated by DFT in the unmodulated phase. (b) Electronic band structure of TaTe$_4$ calculated by DFT in the CDW phase. (c) Comparison between ARPES dispersion and DFT calculations in the unmodulated phase along $\Gamma$-X, showing the predicted hole-like state centered at $\Gamma$ and the electron-like metallic states (indicated by green arrows) not predicted by the calculations. (d) Comparison between ARPES dispersions and DFT calculations in the unmodulated phase along $\Gamma$-Z for both polarizations, showing the dispersion of the hole-like state centered at $\Gamma$, visible only with horizontal polarization, and the electron-like state, visible only with vertical polarization. The quasi-1D metallic state is indicated by blue arrows. The calculated bandstructure in the non-CDW phase was shifted 0.24~eV in order to achieve a better agreement between theory and experiment. All measurements were carried out with a photon energy of 75~eV, and the energy-momentum maps were processed using the curvature method \cite{zhang2011} to enhance the intensity of weak spectral features.}
    \label{FIG3}
\end{figure*}

The electronic band structure of TaTe$_4$ was calculated using Density Functional Theory for two types of atomic structures: the unmodulated phase (1$\times$1$\times$1 -- $P4/mcc$) and the CDW-modulated phase (2$\times$2$\times$3 -- $P4/ncc$). The resulting band structures are shown in Figures \ref{FIG3}(a-b). Although TaTe$_4$ lies in the CDW phase in the temperature range of our measurements, we observed no clear signs of band folding. As a matter of fact, and as it will become clear in the following, the band structure of the unmodulated phase describes more accurately our experimental results.

A comparative analysis between the electronic bands observed by ARPES and those predicted by DFT calculations was performed by superimposing of the calculated dispersions of the non-CDW phase on the energy-momentum maps along $\Gamma$-X and $\Gamma$-Z [Figs. \ref{FIG3}(c) and \ref{FIG3}(d)]. We obtained a very good agreement of theory and experiment after a rigid energy shift of the calculated bandstructure by 0.24~eV to stronger binding energies. All the main features of the experimental band structure can find a counterpart in the theoretical bands with the sole exception of the metallic states discussed in the previous sections and pointed out by arrows in Figures \ref{FIG3}(c) and \ref{FIG3}(d). In particular, along $\Gamma$-X, the hole-like state with a maximum at $-0.2$~eV is captured by theory, while there are experimental traces of the lower-lying theoretical bands. On the other hand, along $\Gamma$-Z, one needs to consider the experimental maps measured using both photon polarizations (LH and LV) in order to observe all bands predicted by DFT. In our experimental configuration, LH photons are sensitive to the aforementioned hole-like band while LV photons to an electron-like band with a minimum at around $-1$~eV. The different dependence of these bands on the photoemission matrix elements is an experimental confirmation of their different orbital origin. Indeed, the hole-like band consists mainly of Te-5$p$ orbitals while the electron-like band of Ta-5$d_{z^2}$ orbitals \cite{guster2022}. We further note that while this electron-like band is predicted to cross the Fermi level, our results suggest that it instead bends down and forms a hole-like maximum around the Z point of the Brillouin zone at approximately $-0.4$~eV: just as the other DFT bands it seems to avoid entering into the region of reduced spectral weight. If this region is related to the CDW as Zwick et al. inferred by comparison to their optical conductivity measurements \cite{zwick1999}, the experimentally observed band maximum along $\Gamma$-Z might be therefore an indirect fingerprint of the CDW. We note, however, that the experimental bands do not match the DFT bandstructure of the CDW-modulated phase [Fig. \ref{FIG3}(b)]. This observation might be surprising, but one has to consider that optical conductivity measurements have concluded that no more than 20-30\% of the charge carriers are involved in the CDW phase \cite{zwick1999}. Last but not least, the fact that the metallic states lying in the region of reduced intensity have no counterpart in bulk calculations questions their attribution to bulk bands and calls for a determination of their dimensionality.

In the following we will discuss in more detail the metallic states crossing the Fermi level by presenting experimental evidence that give new insights on their nature and origin.

\subsection{The origin of the metallic states}
\begin{figure*}
    \centering
    \includegraphics[width=1\textwidth]{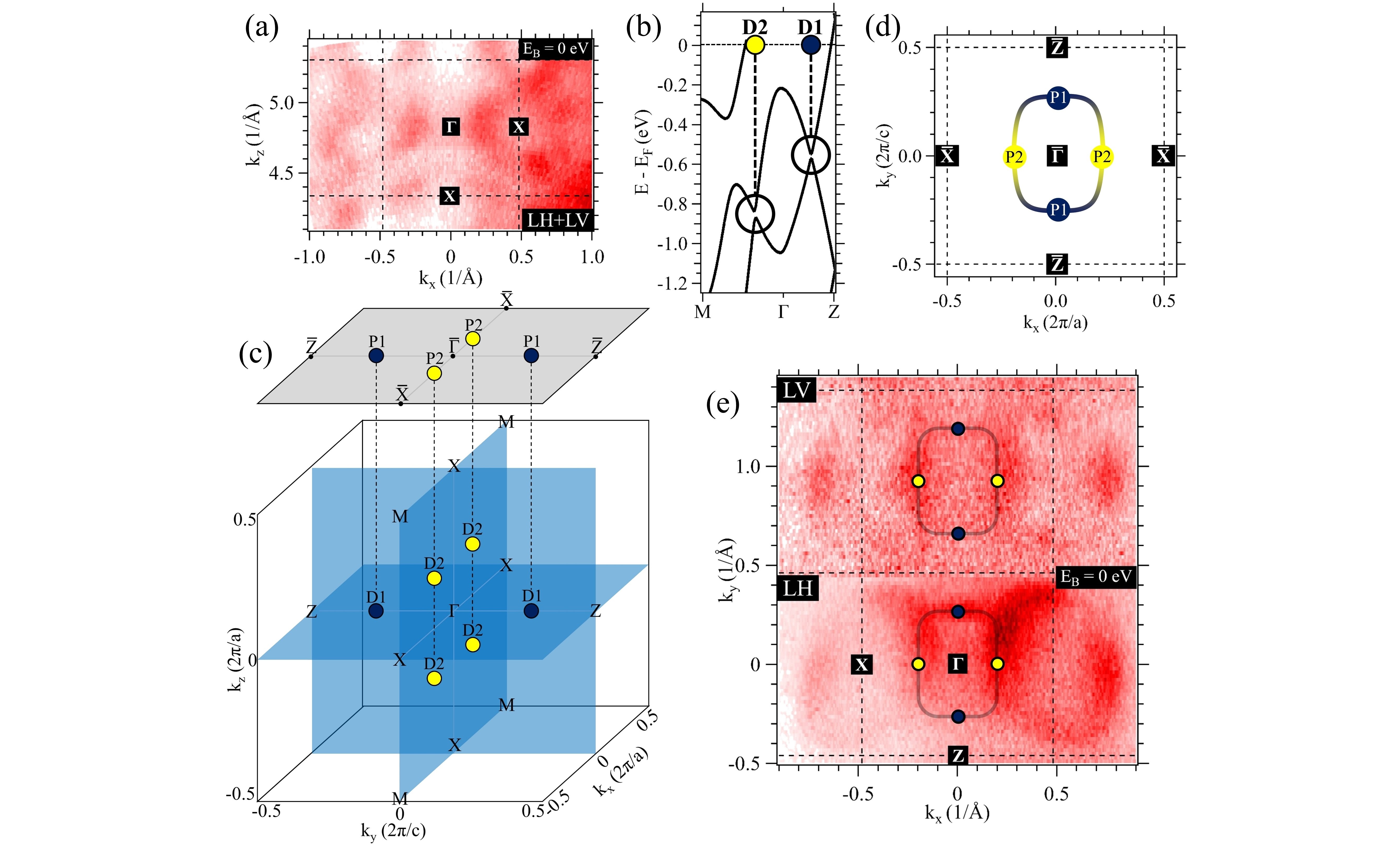}
    \caption{(Color online) (a) ARPES out-of-plane Fermi surface map (E$_B=0$) of TaTe$_4$ obtained by superimposing the maps obtained using photons with both polarizations, showing the out-of-plane dispersion of the electron-like metallic states. The measurements were carried out using photon energies from 50 to 100~eV. (b) Electronic band structure of TaTe$_4$ calculated by DFT along the M-$\Gamma$-Z direction, showing the band crossings along $\Gamma$-Z (blue) and $\Gamma$-M (yellow). The projections of the band crossings at the Fermi level are indicated by points D1 and D2, respectively. (c) Schematic of the positions of the points D1 and D2 in the reciprocal unit cell of TaTe$_4$ and their projections on the surface Brillouin zone indicated by P1 and P2. The indices 1 and 2 refer to a single and double projection, respectively. (d) Schematic of the positions of the points P1 and P2 on the surface Brillouin zone with a qualitative description of the Fermi arcs connecting them. (e) ARPES in-plane Fermi surface maps (E$_B$ = 0) of TaTe$_4$ maps obtained with LH and LV polarization, and the superimposed expected $k$-location of the Fermi arcs. This measurement was carried with a photon energy of 75~eV.}
    \label{FIG4}
\end{figure*}

As discussed previously, there are two sets of metallic states that cross the Fermi level: the states forming the elliptical electron pockets centered at X and those corresponding to the quasi-1D contours normal to $\Gamma$-Z [Fig. \ref{FIG2}(a)]. The former (latter) states are pointed by green (blue) arrows in Figures \ref{FIG1}-\ref{FIG3}. In order to verify the dimensionality of the states forming the elliptical contours, we have checked their out-of-plane dispersion by means of photon energy dependent ARPES. The corresponding out-of-plane Fermi surface map [Fig. \ref{FIG4}(a)] clearly shows that the states are three-dimensional and follow the Brillouin zone periodicity. This experimental observation may come as a surprise for two reasons. First of all, they are not predicted by DFT calculations which should capture all three dimensional bulk-derived states. Secondly, such dimensionality cannot explain the resistivity plateau seen in magnetotransport measurements \cite{luo2017,gao2017}, which is typical of low-dimensional states. In order to get more information on the dimensionality of these states and the spatial extent of their associated wavefunctions, we invite future photoemission studies with more bulk-sensitive probes ($i.e.$ photons in the soft and hard X-ray regimes).

Moving forward, we will discuss the origin of the metallic states behind the quasi-1D Fermi surface contours. As stated previously, in the unmodulated phase, both experiment (Fig. \ref{FIG1}) and calculations (Fig. \ref{FIG3}) agree that the band structure is dominated by a hole-like band at $\Gamma$ having a maximum at $-0.2$~eV. Moreover, there is an electron-like state [see Figs. \ref{FIG3}(a) and \ref{FIG3}(d)] crossing the hole-like band at two points: along $\Gamma$-M and $\Gamma$-Z. As we will shortly explain, this interaction might be at the origin of the metallic quasi-1D states. Specifically, such accidental band crossings could evolve into Dirac points if they result from electronic bands with different symmetries \cite{hasan2010,park2017}. The two bands in question have indeed different orbital symmetries as discussed in the frame of Figure \ref{FIG3}(d). The two alleged Dirac points are shown in Figure \ref{FIG4}(b). It is important here to note that whether or not a gap opening is observed, this could be a consequence of the calculation parameters, more specifically, the number of cells used in the calculations \cite{zhang2010,yazyev2010}. According to Zhang et al. there is a Dirac point along $\Gamma$-Z, while the band crossing along $\Gamma$-M results in a small energy gap \cite{zhang2020}. On the other hand, Guster et al. reported Dirac points in both band crossings \cite{guster2022}.

We now stress the fact that the intersection point between the electron-like and hole-like bands was predicted to occur around $k_y=-0.27$~\r{A}$^{-1}$ along $\Gamma$-Z, which matches perfectly with the $k$-space location of the quasi-1D states. This observation might provide a new explanation for the origin of these states, since it is reminiscent of Fermi arcs associated with bulk Dirac points of 3D topological Dirac semimetals \cite{liu2014,xu2015,liu2014b}.

To examine the plausibility of the latter scenario and understand the dispersion of the quasi-1D states, one must consider all the degeneracy points in the band structure of TaTe$_4$. Figure \ref{FIG4}(a) shows the projection of the Dirac points along $\Gamma$-Z and $\Gamma$-M into the Fermi level, defining the points D1 and D2, respectively. In the reciprocal unit cell, six of these projected points are found: two D1 points at the intersection of the planes $k_z$ = 0 and $k_x$ = 0, referring to the Dirac points along $\Gamma$-Z; and four D2 points located at the plane $k_y$ = 0, referring to the Dirac points along $\Gamma$-M; see Figure \ref{FIG4}(c). The ARPES measurements provide access to the plane $k_z$ = 0, then it is necessary to address the projections of points D1 and D2 into this plane. The Fermi arcs associated with points P1 and P2 are shown schematically in Figure \ref{FIG4}(d) and they are superimposed onto the experimental Fermi surface map in Figure \ref{FIG4}(e). The result shows a decent agreement between the experimentally observed quasi-1D contours and the predicted positions of the Fermi arcs. Despite the fairly good agreement of our experimental findings with the scenario of Fermi arcs, one might be surprised that the bulk Dirac points lie at energies far from the immediate vicinity of the Fermi level, unlike Na$_3$Bi and Cd$_3$As$_2$ \cite{liu2014,xu2015,liu2014b}. On one hand, we stress once again that this is a rather tentative assignment of the quasi-1D states based on our experimental findings. On the other hand, recent calculations by Zhang and coworkers \cite{zhang2020} have provided evidence that TaTe$_4$ should be indeed a topological semimetal making the existence of Fermi arcs highly possible. No matter the origin of the quasi-1D states, we note that their low dimensionality makes them likely candidates for being responsible for the resistivity plateau at low temperatures \cite{luo2017,gao2017}.

As a matter of fact, TaTe$_4$ possesses a unique band structure with metallic states of different dimensionality. Its peculiar Fermi surface consists of textbook 3D states, as well as quasi-1D contours of possible topological origin.

\section{Summary and Conclusions}
In summary, our angle-resolved photoemission spectroscopy experiments on the charge density wave phase of the quasi-1D tetrachalcogenide TaTe$_4$ found that besides the reduction of spectral weight near the Fermi level, there are no clear spectroscopic fingerprints of the CDW itself. On the other hand, our data revealed a well-defined Fermi surface made of various metallic states that had been unnoticed in previous photoemission studies \cite{zwick1999}. Therefore, our work bridges the gap between spectroscopic studies on TaTe$_4$ and magnetotransport work \cite{luo2017,gao2017} that had already reported clear metallic behavior. The metallic states of TaTe$_4$ have different dimensionality characters. Our study hints that the latter low-dimensional states might be attributed to quasi-1D Fermi arcs connecting the projections of band intersections. The potential existence of topological features in a quasi-1D compound makes TaTe$_4$ an interesting material to study the relationship between topology and charge density waves, more precisely, the existence of axions \cite{sekine2021}, as reported in the similar compound (TaSe$_4$)$_2$I \cite{gooth2019,shi2021}.

On completion of this work, a preprint of a new photoemission study on TaTe$_4$ reported a similar band structure to the one in this manuscript \cite{zhang2023}.

\section*{Acknowledgements}
The authors acknowledge SOLEIL and SOLARIS for the provision of synchrotron radiation facilities, proposals No. 20210251 (SOLEIL) and 212021 (SOLARIS). We would like to thank the beamline staff for their support during the experiment in beamline CASSIOPEE (SOLEIL) and URANOS (SOLARIS). Work at ISMO was supported by public grants from the French National Research Agency (ANR), project Fermi-NESt No. ANR-16-CE92-0018, the "Laboratoire d'Excellence Physique Atomes Lumière Matière" (LabEx PALM project MiniVAN) overseen by the ANR as part of the "Investissements d'Avenir" program (reference: ANR-10-LABX-0039), and CNRS International Research Project EXCELSIOR. We thank funding from Ministerio de Ciencias de Colombia, through the grant No. 120480863414. P.H.R. Gonçalves also received financial support from the Brazilian agency CAPES during his stay in ISMO under the process number 88887.370560/2019-00 (2020) and 88887.642783/2021-00 (2022). R.M-P. acknowledges support from CNPq and INCT-Nanocarbono. A.C.G.C. acknowledges the support from the GridUIS-2 experimental testbed, developed under the Universidad Industrial de Santander (SC3-UIS) High Performance and Scientific Computing Centre, with support from UIS Vicerrectoría de Investigación y Extensión (VIE-UIS) and several UIS research groups, as well as other funding resources.

\bibliographystyle{apsrev4-2}
\bibliography{references}

\end{document}